\documentclass[longbibliography,english,reprint,aps,superscriptaddress]{revtex4-2}

\usepackage{babel,calc,amsmath,amsthm,amssymb,graphicx,subfigure}
\usepackage[T1]{fontenc}
\usepackage{mathdots}
\usepackage[unicode=true]{hyperref}
\usepackage{booktabs}
\usepackage{mathtools}
\usepackage{multirow}
\usepackage{lipsum}
\usepackage[normalem]{ulem}
\usepackage[dvipsnames]{xcolor}
\usepackage{appendix}

\definecolor{linkblue}{rgb}{0.2314, 0.4118, 0.6196}
\hypersetup{
     colorlinks=true,       		
     linkcolor=linkblue,          	
     citecolor=linkblue,             
     urlcolor=linkblue,          
 }

\newcommand{\ket}[1]{\left\vert#1\right\rangle}
\newcommand{\bra}[1]{\left\langle#1\right\vert}

\newif\ifdebug

\debugtrue

\ifdebug

\newcommand{\note}[1]{{\color{Orange}{#1}}}
\newcommand\delete{\bgroup\markoverwith{\textcolor{zhliu}{\rule[0.5ex]{2pt}{0.8pt}}}\ULon}

\else

\newcommand{\note}[1]{\ignorespaces}
\newcommand{\delete}[1]{\ignorespaces}

\fi

\begin{document}
\renewcommand{\figurename}{Fig.}

\title                                            {Experimental test of generalized multipartite entropic uncertainty relations}

\author{Zhao-An~Wang}
\thanks{These authors contributed equally to this work.}
\affiliation{CAS Key Laboratory of Quantum Information, University of Science and Technology of China, Hefei 230026, People's Republic of China}
\affiliation{CAS Centre for Excellence in Quantum Information and Quantum Physics, University of Science and Technology of China, Hefei 230026, People's Republic of China}

\author{Bo-Fu~Xie}
\thanks{These authors contributed equally to this work.}
\author{Fei~Ming}
\thanks{These authors contributed equally to this work.}
\affiliation{School of Physics \& Optoelectronic Engineering, Anhui University, Hefei 230601,  People's Republic of China}

\author{Yi-Tao~Wang}
\thanks{These authors contributed equally to this work.}
\affiliation{CAS Key Laboratory of Quantum Information, University of Science and Technology of China, Hefei 230026, People's Republic of China}
\affiliation{CAS Centre for Excellence in Quantum Information and Quantum Physics, University of Science and Technology of China, Hefei 230026, People's Republic of China}

\author{Dong~Wang}
\email{dwang@ahu.edu.cn}
\affiliation{School of Physics \& Optoelectronic Engineering, Anhui University, Hefei 230601,  People's Republic of China}

\author{Yu~Meng}
\email{mengyu23@mail.ustc.edu.cn}

\author{Zheng-Hao~Liu}
\affiliation{CAS Key Laboratory of Quantum Information, University of Science and Technology of China, Hefei 230026, People's Republic of China}
\affiliation{CAS Centre for Excellence in Quantum Information and Quantum Physics, University of Science and Technology of China, Hefei 230026, People's Republic of China}

\author{Jian-Shun~Tang}
\email{tjs@ustc.edu.cn}
\affiliation{CAS Key Laboratory of Quantum Information, University of Science and Technology of China, Hefei 230026, People's Republic of China}
\affiliation{CAS Centre for Excellence in Quantum Information and Quantum Physics, University of Science and Technology of China, Hefei 230026, People's Republic of China}
\affiliation{Hefei National Laboratory, University of Science and Technology of China, Hefei 230088, People's Republic of China}

\author{Liu~Ye}
\affiliation{School of Physics \& Optoelectronic Engineering, Anhui University, Hefei 230601,  People's Republic of China}

\author{Chuan-Feng~Li}
\email{cfli@ustc.edu.cn}

\author{Guang-Can~Guo}
\affiliation{CAS Key Laboratory of Quantum Information, University of Science and Technology of China, Hefei 230026, People's Republic of China}
\affiliation{CAS Centre for Excellence in Quantum Information and Quantum Physics, University of Science and Technology of China, Hefei 230026, People's Republic of China}
\affiliation{Hefei National Laboratory, University of Science and Technology of China, Hefei 230088, People's Republic of China}

\author{Sabre~Kais}
\affiliation{Department of Chemistry, Department of Physics, and Purdue Quantum Science and Engineering Institute, Purdue University, West Lafayette, IN 47907, USA}

\date{\today}

\begin{abstract}
    Entropic uncertainty relation (EUR) formulates the restriction of the inherent uncertainty of quantum mechanics from the information-theoretic perspective.
    A tighter lower bound for uncertainty relations can provide information-theoretic security to quantum communication protocols. Recently, a generalized EUR (GEUR) for the measurement of multiple observables in arbitrary many-body systems has been formulated. Here, we experimentally test this GEUR using a four-photon entangled state with a controllable decoherence channel and show that for the tripartite scenario, the GEUR improves the entropic bound from Renes--Boileau's
    famous results. As an application, we further demonstrate an improvement of the secure key rate in quantum key distribution from the GEUR. Our results extend the test of EURs into multipartite regimes and may find applications in practical quantum cryptography tasks.
\end{abstract}

\maketitle

\textit{Introduction.}---The uncertainty principle, first described by Heisenberg \cite{Heisenberg}, shows the infeasibility of determining the position and momentum of a single particle simultaneously since these two quantities are incompatible in the quantum realm. 
Through decades, uncertainty relations have developed from variance-based ones to entropy uncertainty relations (EUR) for quantum systems \cite{Deutsch}.
EURs were further improved by Kraus \cite{kraus1987}, Massen and Uffink \cite{maassen1988} with a concise expression for the lower bound of two incompatible observables $\hat{R}$ and $\hat{K}$ on a single system:
\begin{align}
    H(\hat{R}) + H(\hat{K}) \ge  - {\log _2}{c(\hat{R};\hat{K})} =: {q_{\rm MU}},
    \label{Eq.singlemu}
\end{align}
 $H(\cdot)$ is the Shannon entropy of probablity distribution of the results of observables, the lower bound is only determined by the complementarity of the observables $\hat{R}$ and $\hat{K}$ with $c(\hat{R};\hat{K}) = \mathop {\max }_{i,j} {\left| {\langle{\varphi _i}|{\phi _j}\rangle } \right|^2}$ ($|\varphi_i\rangle$ and $|\phi_j\rangle$ are eigenstates of the two observables).
 Eq. (\ref{Eq.singlemu}) perfectly addresses state dependence and provides a quantifiable measure of uncertainty in an information-theoretic manner, however, it is limited to single-quantum systems. 
 
Renes \textit{et al.}\,\cite{Renes} and Berta \textit{et al.}\,\cite{Berta} then considered the composite system, where 
the information of system ${\rm A}$ can be detected by a quantum memory system ${\rm B}$ which is correlated with it, hence named as quantum-memory-assisted entropic uncertainty relation (QMA-EUR). 
Then Eq. (\ref{Eq.singlemu}) is reformulated as:  
\begin{align}
    S(\hat{R}\left| {\rm B} \right.) + S(\hat{K}\left| {\rm B} \right.) \ge - {\log _2}{c(\hat{R};\hat{K})}+ S({\rm A}|{\rm B}), 
    \label{Eq.memory} 
    \end{align}
where $S(\hat{R}\left| {\rm B} \right .) = S({\rho _{\hat{R}{\rm B}}}) - S({\rho _{\rm B}})$ denotes the von Neuman entropy of unilateral measurement $\hat{R}$ applied on ${\rm A}$ (conditioned on another correlated system ${\rm B}$) and $S(\rho) = -\text{Tr}(\rho {\log _2}\rho)$ is von Neumann entropy of $\rho$. The corresponding post-measurement state is ${\rho _{\hat{R} \rm B}} = \sum_i (\left| {{\varphi _i}} \right\rangle \left\langle {{\varphi _i}} \right|_{\rm A} \otimes {\mathbb{I}_{\rm B}} )\rho _{\rm AB}(\left| {{\varphi _i}} \right\rangle \left\langle {{\varphi _i}} \right|_{\rm A} \otimes {\mathbb{I}_{\rm B}})$ (the same for $\hat{K}$), where $\mathbb{I}$ is the identity matrix. 
It is noticeable the right-hand side is much tighter due to an additional term --- the information of ${\rm B}$ possessed by ${\rm A}$, conditional entropy S({\rm A}|{\rm B}). 
Although the QMA-EUR was initially developed to enhance the predictive accuracy of the uncertainty principle among quantum systems, it has also become a fundamental tool in quantum information processing, such as building the entanglement criterion\,\cite{Huang,Partovi,Wang22}, certifying quantum randomness\,\cite{Vallone}, deriving quantum steering inequality\,\cite{Schneeloch,Riccardi,Kriv}. Different techniques for constructing useful QMA-EURs in the regime of two-measurement of bipartite systems have been proposed\,\cite{LMHu,JFeng,LMHu2,ZYXu,YLim,MNChen,YBYao}, with several experimental researches implemented in all-optics\,\cite{li2011experimental,prevedel2011experimental,ding2020experimental}, neutron\,\cite{demirel2019experimental}, and diamond\,\cite{ma2016experimental} platforms. The role of EUR in secure quantum communication, particularly in the context of quantum key distribution (QKD), is indispensable\,\cite{Devetak,Christandl,Scarani}, since a tighter bound means a higher quantum-secure-key (QSK) rate and thus higher security\,\cite{Coles1}. 
In building a secure quantum communication network for the future, researchers require not only the simple two-particle or three-particle system version of the EURs but also the more general many-body system version. 
Adabi \textit{et al.}\,\cite{Adabi} introduced the Holevo quantity and mutual information in entropic uncertainty relations to provide an improved lower bound of Eq.\,(\ref{Eq.memory}), which is the tool we employed to build our experiments. 

In this work, we experimentally test the generalized multipartite entropic uncertainty relations. 
We devise a photonic platform that can experimentally generate four-qubit Greenberger–Horne–Zeilinger (GHZ) entanglement states and arbitrary multilateral measurements. Assisted by this platform, we confirm the tightness of the GEUR for a four-partite system. Finally, we introduced an application of the tripartite QMA-EUR by improving the lower bound of the QSK rate by adopting three-qubit Werner states on the same platform.


\bigskip

\textit{GEUR.}---Before we discuss our experiments and main results, we provide some background information. We focus on the generalized EUR (GEUR) reported in Ref.\cite{WangDong}, which starts with a tripartite QMA-EUR bound that is tighter than {Renes--Boileau}'s [Eq.\,(\ref{Eq.memory})], then improves to a generalized version for multi-measurement in the presence of arbitrary multipartite systems.

\begin{figure}[t]
    \centering
    \includegraphics[width=0.45\textwidth]{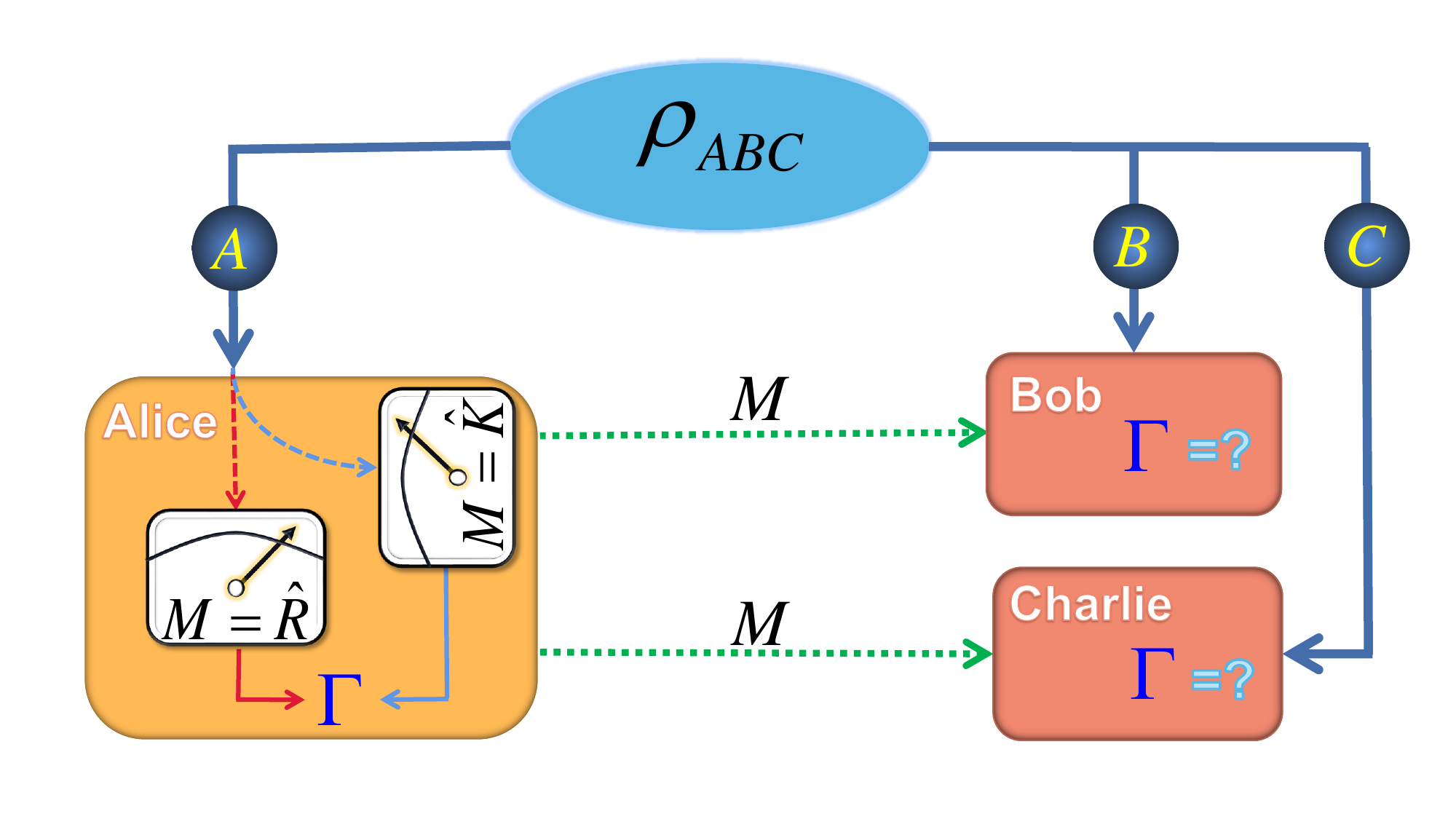}
    \caption{\textbf{Schematic illustrating the uncertainty game in a tripartite quantum system.} A tripartite state $\rho_{\text{ABC}}$ is distributed to Alice, Bob, and Charlie, respectively. Alice first performs either $\hat{R}$ or $\hat{K}$ measurement observables on ${\rm A}$ and she has a measured outcome $\Gamma$, and then announces the measurement choice $M$ to Bob and Charlie. When both Bob and Charlie predict $\Gamma$ with the bound of guess uncertainty $q_{\rm MU}$, they win this game.}
    \label{figure11}
    \end{figure}

The tighter tripartite QMA-EUR is illustrated by an uncertainty game among three correlated subsystems: as can be seen in Fig.\,\ref{figure11}, Alice, Bob, and Charlie, share a tripartite state $\rho_{\rm ABC}$ prepared in advance. Alice randomly chooses one of two observables ($\hat{M}_i= \hat{R},\hat{K}$) and applies it to her system, obtaining the outcome $\Gamma_{\hat{M}_i}$. Alice then notifies her measurement choice $\hat{M}_i$ to Bob and Charlie. If and only if both Bob and Charlie guessed the outcome of Alice's measurement, they would win the game. The winning probability is bounded by the information that Bob and Charlie hold about Alice. This can be calculated with a tripartite entropic uncertainty relation by Renes and Boileau\,\cite{Renes}, and further developed in References \cite{liu,BFXie}, as: 
    \begin{align}
   S( {\hat{R}|{\rm B}} ) + S( {\hat{K}|{\rm C}} ) \ge  {q_{\rm MU}},
   \label{eq:Renes}
    \end{align} 
    the term $q_{\rm MU}$ is as same as in Eq.\,(\ref{Eq.memory}), the core difference is that on the left-hand side, this quantifies the lower bound of the uncertainty of the unilateral measurement on subsystems ${\rm B}$ and ${\rm C}$. 

However, the right-hand side value of Eq.\,(\ref{eq:Renes}), hereafter called Renes--Boileau's bound, is irrelevant to the correlations between the subsystems in the multipartite system; this is in contrast to the case of QMA-EUR, in which the correlation quantified by the conditional entropy is used to improve the bound. Consequently, the Renes--Boileau's bound may not be tight for many multipartite systems with rich and varied forms of entanglement. This is improved in Ref.\cite{WangDong} by taking the effect of quantum correlations into account and improving the bound of entropy, giving an upper bound of the winning probability for the guessing game.

    Formally, the improved tripartite EUR in terms of the entropy, conditional entropy of subsystems, and the Holevo quantities can be stated as:
    \begin{align}
    S( {\hat{R}|{\rm B}}) + S( {\hat{K}|{\rm C}}) \ge {q_{\rm MU}}+\max \left\{ {0,\Delta } \right\},
    \label{eq:theo1}
    \end{align}
    where $\Delta=S({\rm A})-[{\cal{I}}(\hat{R}:{\rm B})+{\cal{I}}(\hat{K}:{\rm C})]$, $S({\rm A}) =  - \sum _i { \lambda _{\rm A}^i } {\log _2} \lambda_{\rm A}^i$ with $\lambda_{\rm A}^i$ being the $i$-th eigenvalue of state $\rho_{\rm A}$ and ${\cal{I}}(\hat R:{\rm B}) = S ({{\rho _{\hat{R}}}})+ S( {{\rho _{\rm B}}}) - S( {{\rho _{\hat{R}{\rm B}}}} )$ is Holevo quantity denoting the upper bound of Bob's accessible information about Alice's measurement results, and for Charlies with ${\cal{I}}(\hat S:{\rm C}) = S( {{\rho _{\hat{K}}}} ) + S( {{\rho _{\rm C}}} ) - S( {{\rho _{\hat{K}{\rm C}}}} )$. The tightness of this bound is dedicated by $\Delta$ on the right-hand side, especially when $\Delta> 0$. When $\Delta\leq 0$, it naturally recovers {Renes--Boileau}'s result.
    
    Technically, compared to tripartite scenarios, a EUR of multiple measurements in multiparty systems is essential in realistic many-body-based quantum information processing, which is deemed as a general case. The GEUR for measurements of multiple observables in a four qubit-system is expressed as 
    \begin{align}
    S( {\hat{R}|{\rm B}} ) + S( {\hat{K}|{\rm C}}) + S(\hat{Q}|{\rm D}) \ge {\cal{B}}_{\rm MU}+\max\{0, \Delta_3\},
    \label{Eq.4qubitgn}
    \end{align}
    where $\Delta_3=\frac{3}{2}S({\rm A})-[{\cal I}(\hat{R}:{\rm B})+{\cal I}(\hat{K}:{\rm C})+{\cal I}(\hat{Q}:D)]$, ${\cal{B}}_{\rm MU}=-[{\log_2}{c(\hat{R};\hat{K})}+{\log_2}{c(\hat{R};\hat{Q})}+{\log_2}{c(\hat{K};\hat{Q})}]/2$. Without losing generality, we choose $\hat{R},\hat{K},\hat{Q}$ to be canonical Pauli observables and thus  ${\cal{B}}_{\rm MU}=3/2$.
            
\begin{figure*}[t]
    \centering
    \includegraphics[width=1.02\textwidth]{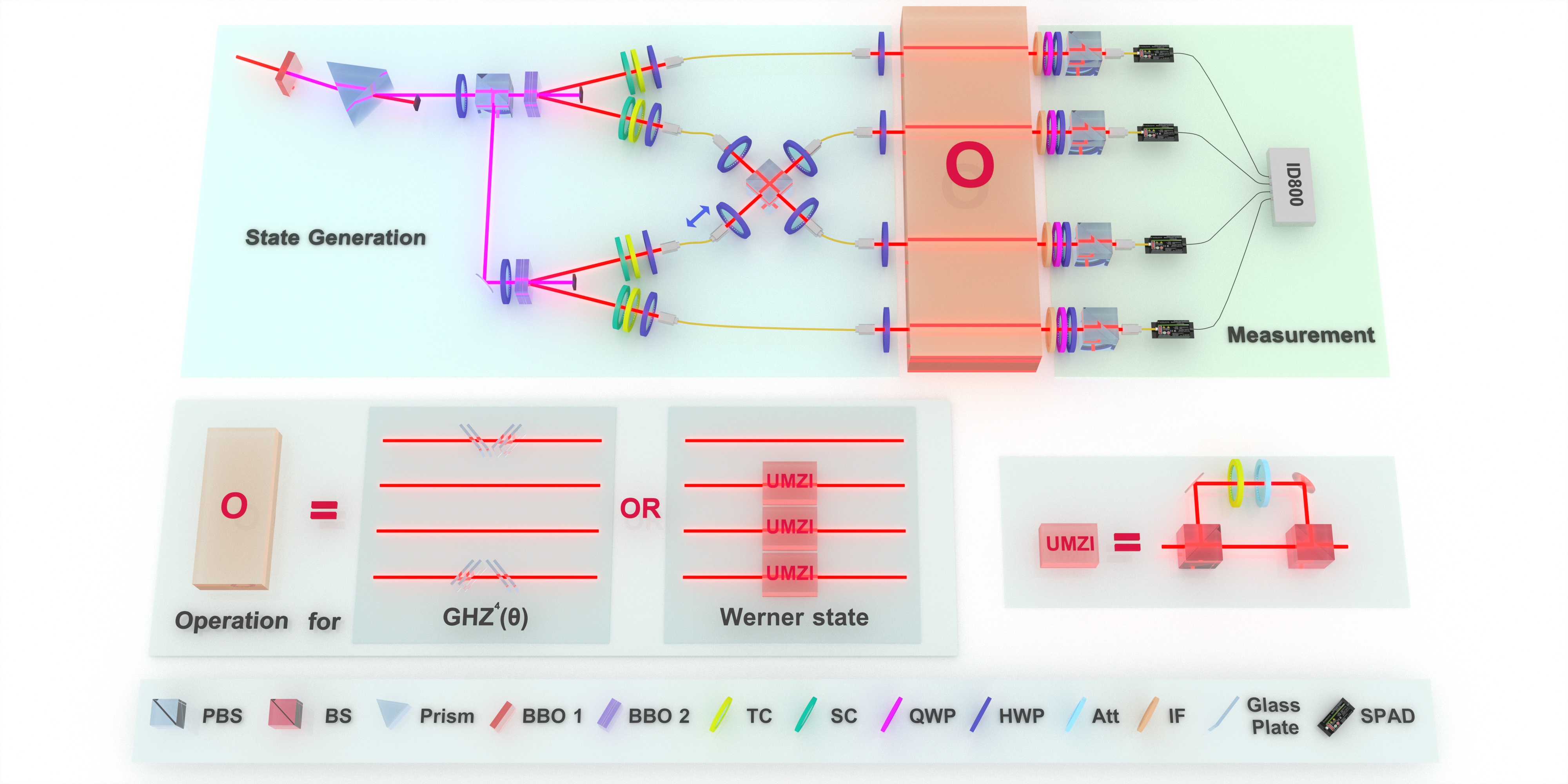}
    \caption{\textbf{Experimental setup for GHZ state.} In the state preparation stage, an ultraviolet laser is generated by a nonlinear crystal (BBO 1) pumped by a pulsed red laser, and the following prism filters out the red laser. Then two pairs of maximally entangled photons were generated by pumping nonlinear crystals (BBO 2) with the ultraviolet lasers. Two photons from each pair coincide on a PBS to generate a four-photon GHZ state conditioned on coincidence counting. The HWPs before and after the single-mode fibers were used to maintain the state polarization. In the evolution stage, denoted by O, the four-photon GHZ state was further transformed to GHZ-like states Eq.\,(\ref{eq:GHZ_theta}) using tilted glass plates or Werner states Eq.\,(\ref{Eq.werner}) using an unbalanced Mach–Zehnder interferometer (UMZI). In the measurement stages, the four photons underwent polarization analysis and coincidence counting. PBS: polarizing beam splitter; BS: beam splitter; TC(SC): temporal (spatial) compensator; QWP: quarter-wave plates; HWP: half-wave plates; Att: attenuator; IF: interference filter; SPAD: single-photon avalanche detector.}
    \label{figure2}
\end{figure*}
    
\textit{Experiment.}---To experimentally demonstrate the tightness of Eq.\,(\ref{Eq.4qubitgn}) and the improvement compared with Renes--Boileau's bound, we prepared 4-qubit GHZ states as the testbed family of a multipartite system from two polarization-entangled photons pairs. The correspondence between the computational bases and photonic polarization states read $\ket{0}\equiv\ket{H},\, \ket{1}\equiv\ket{V}$, with $\ket{H}$ and $\ket{V}$ denoting the horizontal and vertical polarization states of photon, respectively. As depicted in Fig.\,\ref{figure2}, two beam-like entangled photon pairs centered at the wavelength of 808 nm were generated via the spontaneous parametric down-conversion process by pumping two type-II $\beta$-barium borate (BBO 2) crystals sandwiches \cite{zhang2015experimental}. The pumping laser is centered at 404 nm, generated through the frequency doubling process of another BBO crystal (BBO 1) pumped by a femtosecond pulsed laser centered at 808 nm with a repetition rate of 76 MHz. The prism after BBO 1 filters out the 808 nm pumping laser. After compensating for the spatial and temporal walkoff by the birefringent crystals, two-qubit entangled source were prepared in the Bell state $\ket{\Phi^+}=(\ket{HH}+\ket{VV})/\sqrt{2}$ with fidelity of approximate 99.0\%. 
The polarization states of the photons were maintained by six pairs of half-wave plates before and after the fiber to revert the unwanted polarization rotation induced by the fibers.
   
\begin{figure}[t]
    \centering
    \includegraphics[width=8.5cm]{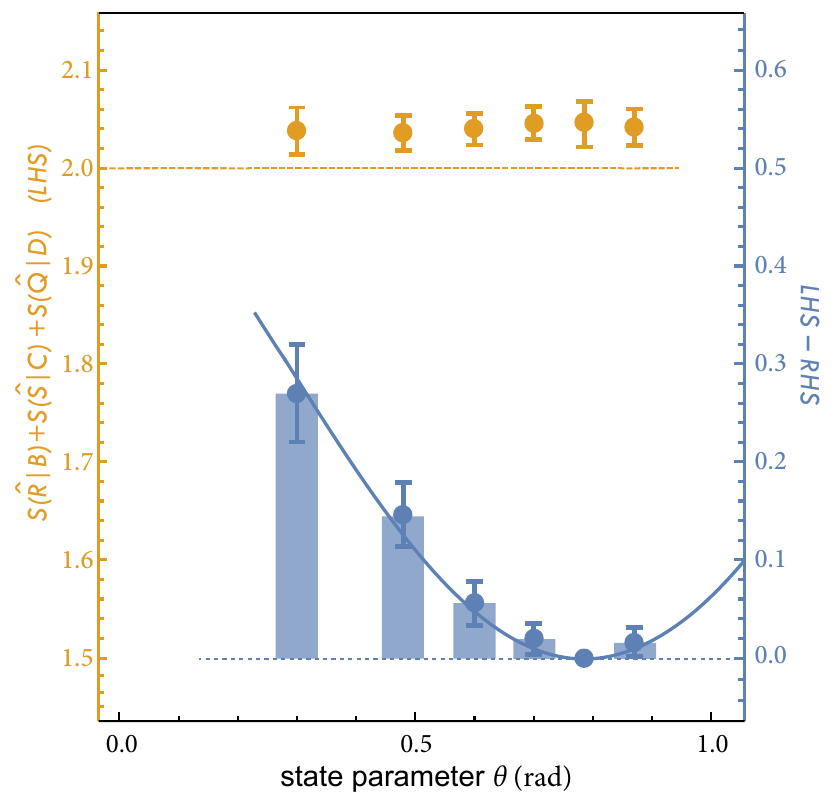}
    \caption{\textbf{Experimental certification of the tightness of the improved bound.} The yellow dots are the experimental value of the uncertainty [left-hand side (LHS) of Eq.(\ref{eq:theo1})], i.e., $S(\hat{R}\left| {\rm B} \right.) + S(\hat{K}\left| {\rm C} \right.)+ S(\hat{Q}\left| {\rm D} \right.)$, which is identically equal to constant 2. The blue points presenting the difference between the uncertainty and the improved bound [right-hand side (RHS) of Eq.(\ref{eq:theo1})] together with the corresponding histograms indicate the tightness of the improved bound varying with $\theta$. The tightest condition is achieved given the state parameter $\theta=\pi/4$, and all the observed relations imply an improvement of the Renes–Boileau's bound where the difference between the uncertainty and ${\cal{B}}_{\rm MU}$ is identically equal to ${\frac{1}{2}}$. The error bars denoting the $1\sigma$ standard deviation are deduced by the Monte Carlo simulation with the detected photon counts obeying Poisson distribution.} 
    \label{figure3}
\end{figure}

\begin{figure}[t]
    \centering
    \includegraphics[width=8.5cm]{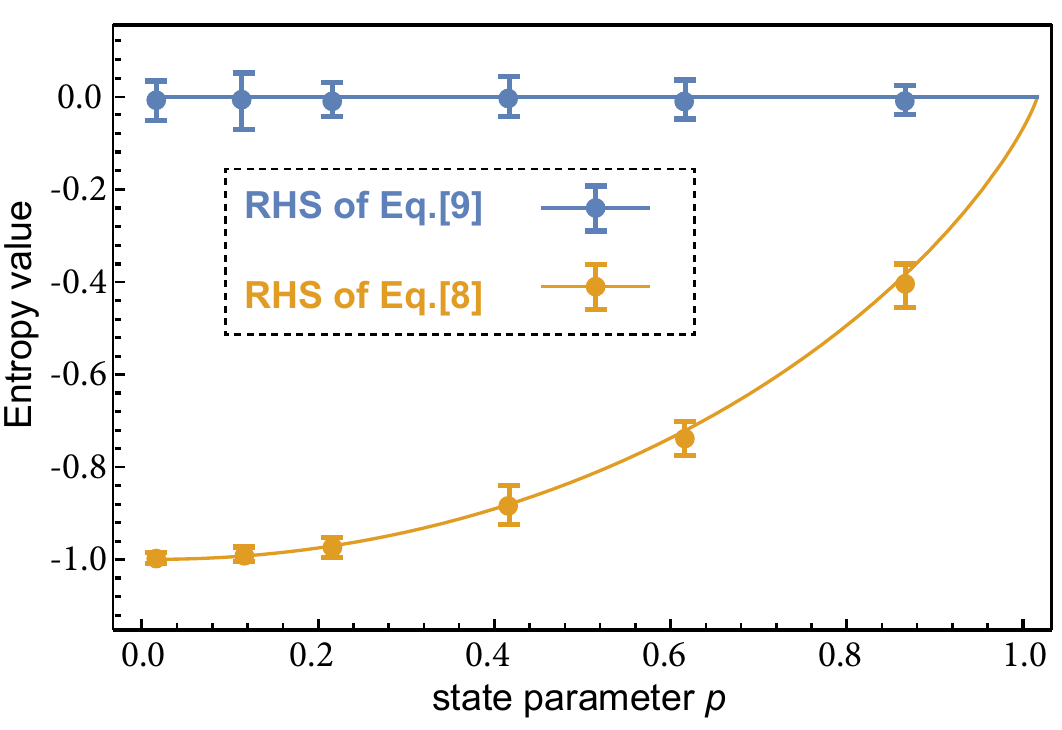}
    \caption{\textbf{Improved lower bound of QSK rate.} The blue dots are the experimental results of the QSK rate of different Werner states (right-hand side of Eq.\,(\ref{Eq.KLBbinary})), whereas the yellow dots (consistent with the theoretical yellow lines) represent the experimental results of the right-hand side of Eq.\,(\ref{Eq.QKDbinary}). The error bars denoting the $1\sigma$ standard deviation are deduced by the Monte Carlo simulation with the detected photon counts obeying Poisson distribution.} 
    \label{figure4}
\end{figure}
    
    To prepare the 4-qubit GHZ entangled state, we took one photon from each of the entangled photon pairs to make the two photons coincide on a polarizing beam splitter (PBS). The PBS transmits $\ket{H}$-polarized photons and reflects $\ket{V}$-polarized photons, respectively, so only two incident photons with the same polarization can be output coincidentally from different ports. We selected the events via four-fold coincidence photon counting detected at each output port simultaneously, and the PBS acted as a parity operator $\ket{HH}\bra{HH}+\ket{VV}\bra{VV}$\,\cite{Bodiya06}. The successful postselection projected the output state onto the four-qubit GHZ state:
    \begin{align}
    \left|\rm GHZ^4\right\rangle=\frac{1}{\sqrt{2}}(|H H H H\rangle+|V V V V \rangle)_{\rm ABCD},
    \label{Eq.ghz}
    \end{align}
    with a probability of $\frac{1}{2}$. We have adjusted carefully the length of optical paths before the PBS to ensure the maximal multiple-photon interference visibility and employed a tiltable quarter-wave plate (QWP) oriented at $0^\circ$ at one photon path to compensate the relative phase between the two terms of Eq.\,(\ref{Eq.ghz}). The photons in the four paths were filtered by 3-nm band-pass filters before measurement. Furthermore, we expanded the family of prepared states by inserting groups of Brewster windows to adjust the relative amplitude of $\ket{H}$ and $\ket{V}$-polarized wavefunction, which produces the set of experimentally amenable states with different parameters of $\theta$, after a re-normalization, to become:
    \begin{align}
    \ket{{\rm GHZ}^4 (\theta)}=\cos\theta \ket{HHHH}+\sin\theta \ket{VVVV}.
    \label{eq:GHZ_theta}
    \end{align}
    Finally, we used four groups of polarization analysis devices, consisting of a QWP, an HWP, and a PBS, to perform complete sets of projective measurements.

    We used quantum state tomography on various subsystems to deduce their density matrices with maximum-likelihood estimation \cite{James01} and the corresponding von Neumann entropies. In particular, the Holevo quantities were calculated after the measurement of a specific observable has taken place and the results for testing the multipartite EUR are illustrated in Fig.\,\ref{figure3}. Theoretically, both the uncertainty (left-hand side value of Eq.\,(\ref{Eq.4qubitgn})) and the Renes--Boileau's bound ${\cal{B}}_{\rm MU}$ remain independent of the state parameter, the values are 2 and ${\frac{3}{2}}$, respectively. We see that Renes--Boileau's bound is not tight anywhere for this testbed family of states. Experimentally, we measured the sum of conditional entropies for calculating the uncertainty, $S(\hat{R}\left| {\rm B} \right.) + S(\hat{K}\left| {\rm C} \right.)+ S(\hat{Q}\left| {\rm D} \right.)$ as all close to but slightly above 2. This difference is due to the slight deviation of our experimental state from the ideal state given by Eq.\,(\ref{eq:GHZ_theta}).  
    However, in the multipartite EUR, the difference between the uncertainty and the new bound can be reduced down to near 0 when $\theta=\pi/4$, which significantly improves the situation. Besides, we tested the looser part of the improved bond to fulfill the improvement attributed to the non-vanishing quantity $\Delta_3$. The improvement declines to zero when $\theta$ varies to 0 theoretically, and we show the results down to $\theta=0.3 (rad)$ in the experiment limited by the extinction ratio of the Brewster window array.
    

\textit{Application in quantum key distribution.---}The tighter bound of uncertainty should be of fundamental relevance to the security analysis of QKD protocols in practical many-body systems. Particularly, we focus on tripartite systems. Regarding two honest participants (say, Alice and Bob), who are sharing a key by communicating over a public channel in the QKD processing, the key is secret to any third-party eavesdropper (e.g., Dave). In the eavesdropper-existing communication process, Dave (eavesdropper) prepares a quantum state $\rho_{\rm ABD}$; and sends subsystems ${\rm A}$ and ${\rm B}$ to Alice and Bob respectively while keeping ${\rm D}$. By connecting Renes–Boileau's bound and considering $S( {\hat{R}|{\rm B}} )\leq S( {\hat{R}|\hat{R}'} )$ and $S( {\hat{K}|{\rm B}} )\leq S( {\hat{K}|\hat{K}'} )$, the quantity of key $K$ that can be extracted by Alice and Bob is lower bounded by \cite{li2011experimental}
\begin{align}
K\geq q_{\rm MU}-S( {\hat{R}|\hat{R}'} )-S( {\hat{K}|\hat{K}'} ).
\label{Eq.QKDbinary}
\end{align}
Exploiting Eq. (\ref{eq:theo1}), the improved QSK rate is expressed by
\begin{align}
K_{\rm LB} \ge {q_{\rm MU}} +  \max \left\{ {0,\Delta } \right\} - S\left( {\hat R|\hat{ R}'} \right) - S\left( {\hat K|\hat{K}'} \right).
\label{Eq.KLBbinary}
\end{align}
Tighter than the QSK rate attained by Berta {\it et al}, this relation is capable of efficiently improving the security of QKD protocols.
 
    To demonstrate the improvement in QSK rate in the current scenarios, we resorted to a class of Werner states. For simplicity, we experimentally prepared the system state as a three-photon Werner state, which is expressed by:
    \begin{align}
    \rho_{\rm W}(p)=p\ket{\text{GHZ}^{3}}\bra{\text{GHZ}^{3}}+\frac{(1-p)}{8}\mathbb{I},
\label{Eq.werner}
\end{align}
    where $\ket{\text{GHZ}^{3}}=(\ket{HHH}+\ket{VVV})/\sqrt{2}$ denotes the three-qubit GHZ state and $\frac{\mathbb{I}}{8}$ is the maximally mixed state. To prepare the Werner state, we replaced the Brewster windows in each path with an unbalanced Mach--Zehnder interferometer (UMZI) in the experimental setup (see Fig.\,\ref{figure2}) and projected the photon ${\rm C}$ to a fixed basis $(\ket{H}+\ket{V})/\sqrt{2}$. The variable attenuator in the interferometer allowed to adjust the mixing parameter $p$. The length difference between the two arms of each UMZI was roughly 0.8 m, and the coincidence interval was set to 0.8 ns to reject all events where some photons propagated through the long arm and the others through the short arm.
    
    Fig.\,\ref{figure4} shows the improvement of the QSK rate's lower bound with bilateral measurements. The scenario can be described as the subsystems ${\rm A}$, ${\rm B}$ are sent to Alice and Bob respectively, and the eavesdropper (Dave) keeps ${\rm D}$. For simplicity, we limit both Alice and Bob to employ ${\sigma_y}$ and ${\sigma_z}$ measurements as the observables respectively. The new bound exploiting multipartite EUR surpasses the best previous result; in the limit of $p\to0$ the improvement becomes as large as 1. The experimental results, in good agreement with the theoretical prediction, clearly demonstrated the potential of our proposed EUR in implementing more efficient practical QKD using the existing infrastructure. 
    \vspace{10pt}
    

\textit{Discussion.---}In this work, we experimentally demonstrate a Generalized Entropic Uncertainty Relation (GEUR) that applies to measurements of multiple observables under a multipartite architecture. We focused on canonical Pauli measurements, which are known to be incompatible and challenging to measure experimentally within four-qubit GHZ states. 
This multipartite experimental layout not only provides valuable insights but also serves as a heuristic strategy for conducting entropy-based information experiments involving more complex and higher-dimensional measurements.

Moreover, we emphasize the practical implications of our work in the context of quantum key distribution (QKD). By applying the GEUR, a more stringent bound for the QKD secret key rate can be established. This breakthrough has significant implications for enhancing the security of prospective quantum communication networks. We believe that our observations offer profound insights into entropy-based uncertainty relations for multi-measurement scenarios, particularly in the presence of arbitrary multipartite systems. This advancement is thought to be vital in boosting the security of realistic QKD networks. 
Looking ahead, the application of GEUR in various fields such as quantum cryptography, quantum randomness, quantum steering, and quantum metrology holds great promise. These areas can benefit from our findings and pave the way for future research directions in quantum information sciences. Overall, our study represents a crucial step towards harnessing the potential of the GEUR in practical applications, opening up new avenues for advancements in quantum technology.

\begin{acknowledgements}
\vspace{2pt}
    This work is supported by the Innovation Program for Quantum Science and Technology (No. 2021ZD0301200), the National Natural Science Foundation of China (Nos. 12174370, 12174376, and 11821404), the Youth Innovation Promotion Association of Chinese Academy of Sciences (No. 2017492), the Fok Ying-Tung Education Foundation (No. 171007), the Open Research Projects of Zhejiang Lab (No. 2021MBOAB02). D.W. acknowledges the support by the National Natural Science Foundation of China (Grant Nos. 12075001, 61601002 and 12175001), Key Research and Development Project of Anhui Province (Grant No. 2022b13020004), and the fund from CAS Key Laboratory of Quantum Information (Grant No. KQI201701). S.K. acknowledges the financial support by the National Science Foundation under award number 1955907.
\end{acknowledgements}

\providecommand{\noopsort}[1]{}\providecommand{\singleletter}[1]{#1}%

\end{document}